\documentclass[10pt]{article}
\usepackage{url,amsmath,amsthm,amsfonts,amssymb}
\usepackage{comment,color}

\newcommand{\thmabove}{7.5pt}
\newcommand{\thmbelow}{0pt}

\newtheoremstyle{mythmstyle}
  {\thmabove}   % Space above
  {\thmbelow}   % Space below
  {}            % Font of theorem body (e.g., \itshape)
  {}            % Indent amount (empty = no indent, \parindent = para indent)
  {\bfseries}   % Thm head font
  {. }          % Punctuation after thm head
  {2.5pt}       % Space after thm head (\newline = linebreak)
  {\thmname{#1}\thmnumber{ #2}\thmnote{ \normalfont (#3)}}   % Thm head spec
\theoremstyle{mythmstyle}
\newtheorem{theorem}{Theorem}[section]\numberwithin{equation}{section}

\newtheorem{definition}[theorem]{Definition}

\newcommand{\EquationName}[1]{\label{eq:#1}}

{\left(\begin{smallmatrix}}
{\end{smallmatrix}\right)}

\title{Submodular welfare maximization}
\author{Samira Samadi}
\date{}

\begin{document}
\maketitle

%%%%%%%%%%%%%%%%%%%%%%%%%%%%%%%%%%%%%%%%%%%%%%%%%%%%%%%%%%%%%

\section{Introduction}

% ===========================================================
% the following paragraph is added to introduce a problem and also to justify
% the definition of valuations as function on subsets of items.

We study different variants of the welfare maximization problem in combinatorial
auctions. Assume there are $n$ buyers interested
in $m$ items. Indeed they have different valuations on each item, i.e.\ buyer
$i$ has a valuation of $w_i(j)$ on item $j$. In the simplest model, a
collection of items $S$ worths $\sum_{j \in S} w_i(j)$ for buyer $i$; However,
in practice, collection of items might have different values than the sum of
values of individual items. Therefore, in a more accurate model, we should take
this into account and define the valuation of each buyer as a real valued
function on the family of subsets of items, i.e.\ $w_i : 2^{[m]}
\rightarrow \mathbb{R}_{+}$. Our goal is to partition these items among buyers
so as to maximize the total welfare. More precisely, we partition the set of
items into subsets $S_1, \dots, S_n$ and give bundle $S_i$ to buyer
$i$ in order to maximize $\sum_{i=1}^n w_i(S_i)$.

% ===========================================================
% oracle types
\begin{comment}
Unless we consider a 
special class of utility functions with a polynomial-size representation, we
typically resort to an oracle model. An oracle answers a certain type of queries
about a utility function. Two types of oracles have been commonly considered:
\begin{itemize}
\item Value oracle. The most basic query is: What is the value of $w_i(S)$? An
oracle answering such queries is called a value oracle.
\item Demand oracle. Sometimes, a more powerful oracle is considered, which can
answer queries of the following type: Given an assignment of prices to items
$p:[m]\rightarrow \mathbb{R}$, which set $S$ maximizes $w_i(S)
-\sum\limits_{j\in S} p_j$? Such an
oracle is called a demand oracle.
\end{itemize}
\end{comment}
Based on the above discussion, not only we need the valuation of each person on
each item, but also we need to present the valuations on subsets of items
which requires $O(n 2^m)$ bits of data to be stored as the input of the
algorithm. This huge amount of data  is exponential in the number of items,
which could make practical difficulties especially when the number of items,
$m$, is very large. Hence, instead of presenting the whole data to the
algorithm, we employ an oracle who answers queries when asked.
Usually, two types of oracles have been considered:
\begin{description}
 \item[Value oracle] this type of oracle answers the questions of the type
 ``What is the value of $w_i(S)$ for a person $1\leq i \leq n$ and a
collection of items $S$?'' This is obviously the simplest type of oracle who
just looks at the database of valuations with zero computational overload.
\item[Demand oracle] unlike value oracle which does not compute anything, in
this case we have access to a more powerful oracle which has computational
power to solve a maximization problem. In fact, we have access to a powerful
black box which can solve the following problem for us in zero time: assume
$p:[m]\rightarrow \mathbb{R}$ is a price function on items, find the set $S$
which maximizes $w_i(S) - \sum_{j \in S} p_j$.
\end{description}

% ==============================================================================
% the problem is difficult in general, we take special cases
\begin{comment}
 Our goal is to find disjoint sets $S_1,\cdots,S_n$ maximizing the total welfare
$\sum\limits_{i=1}^{n}w_i(Si)$. Regardless of what type of oracle we consider,
there are computational issues that make the problem very hard in general. In
particular, consider players who are single-minded in the sense that each
desires one particular set $T_i$, $w_i(S) = 1$ if $T_i \subseteq S$ and $0$
otherwise. Then both value and demand queries are easy to answer; however, the
problem is equivalent to set packing which is known to have no
$m^{-1/2+\epsilon}$-approximation unless P = NP [13, 27]. Thus, restricted
classes of utility functions need to be considered if one intends to obtain
non-trivial positive results.
A class of particular interest is the class of monotone submodular functions. 
An equivalent definition of this class is as follows: Let  $f_S(j)=f(S+j)-f(S)$
denote the marginal value of item $j$ with respect to $S$. (We write $S + j$
instead of $S \cup \{j\}$ to simplify notation.) Then $f$ is monotone if $f_S(j)
\geq 0$ and $f$ is submodular if $f_S (j) \geq f_T (j)$ whenever $S \subset T$.
This can be interpreted as the property of diminishing returns known in
economics and arising naturally in certain settings. This leads to what we call
the Submodular Welfare Problem.
\end{comment}
Even if we employ the powerful demand oracle, the problem could be very hard to
compute in general. For an example, consider the scenario in which each buyer is
only interested in an specific subset of items, denoted by $T_i$ for buyer $i$.
Therefore he is satisfied when he collects all the items in $T_i$, paying no
attention to additional items, i.e.\ $w_i(S) = 1$ if $T_i \subseteq S$ and $0$
otherwise. In this case, both value and demand queries are trivial, but the
problem is equivalent to the set packing problem which has no
$m^{-1/2+\epsilon}$-approximation unless $P = NP$ \cite{13, 27}. Based
on this
discussion, we restrict the set of permissible utility functions to obtain
non-trivial positive results. In particular, we add two assumptions which are
reasonable in practice.

The first assumption is monotonicity. We expect a person to be more satisfied
when he obtains a larger set of items, i.e.\ we do not have negative valuations
or ``disgusting'' items. This translates to monotonicity of functions $w_i$:
\begin{definition}
A function $f:2^X\rightarrow \mathbb{R}$ is monotone if $f(S) \leq f(T)$
whenever $S \subseteq T$.
\end{definition}
The other assumption is related to the improvement in gains by giving extra
items to players. In practice, a reach person is less excited by obtaining a
Porsche than a relatively poor person. Therefore, it is reasonable to assume
that adding an extra item to a large set is less attractive than adding it to
a relatively smaller one. This translates to submodularity of functions. A
function $f$ defied on the family of subsets of a set is called submodular if
\[
 f(S \cup \{j\}) \geq f(T \cup \{j\}) \qquad S \subset T.
\]
It could be shown that this definition of submodular functions is equivalent to
the following:
\begin{definition}
A discrete function $f:2^{X}\rightarrow \mathbb{R}$ is submodular if $f(S\cup
T)+ f(S\cap T) \leq f(S)+f(T)$. Usually we also assume that $f(\emptyset)=0$.
\end{definition}
Adding these two assumptions leads us to Submodular Welfare Problem.
Conclusively,
\begin{definition} [Submodular welfare maximization]
$m$ buyers and $n$ items are given. 
Each buyer has a monotone submodular valuation $w_i : 2^{[m]} \rightarrow
\mathbb{R}_+$ which is his interest in different subsets of items . The goal is
to partition items into disjoint sets $S_1, S_2,\cdots,S_n$ in order to maximize
$\sum\limits_{i=1}^{n}w_i(S_i)$.
\end{definition}

% ============================================================================
% other areas in which Sub... functions appear, min/max difficulty
\begin{comment}
"Submodular functions arise naturally in combinatorial optimization, e.g. as
rank functions of matroids, in covering problems, graph cut problems and
facility location problems [5, 18, 24]. Unlike minimization of submodular
functions which can be done in polynomial time [10, 23], problems involving
submodular maximization are typically NP-hard. Research on problems involving
maximization of monotone submodular functions dates back to the work of
Nemhauser, Wolsey and Fisher in the 1970Õs [20, 21, 22]."
\end{comment}
Submodular functions appear in other areas like rank functions of matroids, in
covering problems, graph cut problems and
facility location problems \cite{5, 18, 24}. It could be shown that
minimization of
submodular functions could be done in polynomial time (\cite{10, 23}), however
maximization of such functions is typically NP-hard. First studies on
maximization of monotone submodular functions is due to Nemhauser, Wolsey and
Fisher in the 1970Õs \cite{NWF78, FNW78, NW78}.

\begin{comment}
"Recently, there has been renewed interest in submodular maximization due to
applications in the area of combinatorial auctions. In a combinatorial auction,
$n$ players compete for m items which might have different value for different
players, and also depending on the particular combination of items allocated to
a given player. In full generality, this is expressed by the notion of utility
function $w_i : 2^{[m]}\rightarrow R_+$ which assigns a value to each set of
items potentially allocated to player $i$. Since this is an amount of
information exponential in $m$, we have to clarify how the utility functions are
accessible to an algorithm. "
\end{comment}

%%%%%%%%%%%%%%%%%%%%%%%%%%%%%%%%%%%%%%%%%%%%%%%%%%%%%%%%%%%%%%

%%%%%%%%%%%%%%%%%%%%%%%%%%%%%%%%%%%%%%%%%%%%%%%%%%%%%%%%%%
\section{Algorithmic results}

\begin{comment}
"Variantsof the problem arise by considering different classes of valuation
functions $w_i$ and different models (adversarial/stochastic) for the arrival
ordering of the items." [MIJ13]
\end{comment}
We can categorize welfare maximization problem based on two parameters: the
first one is different classes of valuation functions, $w_i$. Usually
assumptions on $w_i$ are monotonicity and submodularity. The other parameter is
the order in which items arrive. We may assume that the division process takes
place when all the items and valuations are known (offline) items arrive one by
one and valuations are known for so far received items and each item should be
assigned upon arrival (online). Results on offline and online welfare
maximization are given in the following. It should be noted that we do not seek
the exact solution, instead an approximation of the solution suffices, i.e.\ a
solution that is within a coefficient of the answer is acceptable.
%\begin{itemize}
%\item\textbf{Offline model}: 
\subsection{Offline Model}
In offline model, we should assign each item to a player after receiving the
whole set of items. In \cite{V08}, a randomized 	continuous
greedy algorithm for
the submodular welfare problem is derived which is a $(1-1/e)-$ approximation.
Interestingly, in the special case when the valuations of players are
identical, the optimal answer is obtained by uniform random solution. It is
shown using information theoretic lower bounds that solving the problem more
accurately (with better approximation factor), an exponential number of value
queries is necessary. Furthermore, in this paper the problem is analyzed for
the two classes \emph{subadditive} and \emph{superadditive} valuation
functions. A set function $f$ is said to be subadditive iff $f(S) + f(T) \geq
f(S \cup T)$ and is said to be superadditive iff for disjoint sets $S$ and
$T$, $f(S) + f(T) \leq f(S\cup T)$. Note that subadditivity is similar to
submodularity, however, submodularity is a stronger condition. It is shown that
approximation factors $\frac{1}{\sqrt{m}}$ and $\frac{\sqrt{\log m}}{m}$
respectively for subadditive and superadditive valuations are the best
approximation factors which can be obtained by asking a polynomial number of
value queries, i.e.\ better approximation factors require asking super
polynomial number of queries. This shows that the above approximation factors
are the best possible ones.

\subsection{Online Model}
\begin{definition}
$m$ items are arriving online, 
and each item should be allocated upon arrival to one of $n$ agents whose
interest in different subsets of items is expressed by valuation functions $w_i
: 2^{[m]} \rightarrow \mathbb{R}_+$. 
Also it is assumed that we only know the agents' valuations on items arrived so
far.
The goal is to maximize
$\sum\limits_{i=1}^{n} w_i(Si)$ where $S_i$ is the set of items allocated to
agent $i$.
\end{definition} 

%\begin{itemize}
%\item 
Since we should decide immediately  upon arrival which person to give the
current item, the simplest framework is that we give the current item to the
person who gets excited the most, i.e.\ we assign the arrived item to the
player whose welfare increases the most. Fisher, Nemhauser and Wolsey who
worked on problems involving maximization of submodular functions, introduced
this greedy algorithm \cite{NWF78, FNW78, NW78}. This greedy strategy is shown
to be
$1/2$-competitive when valuation functions are monotone and submodular.

\begin{comment}
Greedy is a $\frac{1}{2}$ approximationalgorithm. [NWF78]:
The baseline algorithm in this setting is the greedy algorithm,
due to Fisher, Nemhauser and Wolsey, who initiated the study of problems
involving maximization of submodular functions [NWF78, FNW78, NW78]. The greedy algorithm simply allocates each incoming item to the agent who gains fro
 it the most and is 1/2-competitive whenever the valuation functions of the agents ar
 monotone submodular [FNW78, LLN06].
\end{comment}

\begin{comment}
a seminal paper of Karp, Vazirani and Vazirani [KVV90] on
online bipartite matching. This can be viewed as a welfare maximization problem
where one side of the bipartite graph represents agents and the other side
items; each agent i is interested in the items $N(i)$ joined to i by an edge,
and he is completely satisfied by 1 item, meaning the valuation function can be
written as $w_i(S ) =\min{\lvert S \cap N(i) \rvert , 1}$ (a very special case
of a submodular function). Karp, Vazirani and Vazirani gave an elegant
$(1-1/e)$-competitive randomized algorithm, which improves a greedy
1/2-approximation and is optimal in this setting.
\end{comment}
An special case of this problem is reduced to online bipartite matching analyzed
by Karp, Vazirani and Vazirani \cite{KVV90}. In the online bipartite matching
problem, one vertex of the first part of the graph as well as its connections
is given at a time, and upon receiving this information, we should decide which
vertex of the second part to match to it so as to maximize the size of the
matching at the end of the day. The following restriction on valuation
functions reduces online welfare maximization to online bipartite matching;
Assume each agent $i$ is completely satisfied by only one item in an specific
set $N(i)$, i.e.\ $w_i(S) = \min \{ |S\cap N(i)|, 1\}$. Having this assumption,
we form a bipartite graph in which the first part and second parts are
representatives of the agents and the items respectively. Connect an agent $i$
to his $N(i)$ interested items. Then online welfare maximization reduced to
online bipartite matching, for which there exists an elegant
$(1-1/e)$-competitive randomized algorithm in \cite{KVV90}. Note that this is
an
improvement on the greedy $1/2$-competitive solution.
%\item Better algorithms for spacial cases?

%\end{itemize
%\end{itemize}
 
 \section{Truthful Mechanism Design}

So far we have assumed that the valuations of players are known to us when
allocating the items. However, in practice, we might not have access to actual
valuations. In this setting the aim of the mechanism designer is to design a
computationally efficient mechanism in which he hopes that the agents are
truthful and that achieves with an approximation factor the optimal solution
found by the former version of the
problem in which all the information is provided in advance. Now  we give a
formal definition of what it means for mechanism designer to hope that agents
are incentive compatible (\cite{DRY11}):

\begin{definition}
A mechanism with allocation and payment rules $\mathcal{A}$ and $p$ is
\textit{truthfull-in-expectation} if every player always maximizes it's expected
payoff by truthfully reporting it's valuation function meaning that 
\begin{equation}
\EquationName{truthful}
\mathbb{E}[v_i(\mathcal{A}(v))-p_i(v)] \geq
\mathbb{E}[v_i(\mathcal{A}(v'_i,v_{-i}))-p_i(v'_i,v_{-i})]
\end{equation}
for every player $i$, (true) valuation function $v_i$, (reported) valuation
function $v'_i$, and (reported) valuation functions $v_{-i}$ of the other
players. The expectation in \eqref{eq:truthful} is over the coin flips of the
mechanism.
\end{definition}
In \cite{DRY11}, a $(1-1/e)$-approximation truthful in expectation mechanism for
coverage valuations is derived. Also, it is shown, assuming $P!=NP$, even for
known and explicitly given coverage valuations, the approximation factor could
not be improved. However, for submodular valuations, no truthful-in-expectation 
mechanism exists \cite{DV11}.

It is worth noting that without incentive-compatibility constraints, the
welfare maximization problem with submodular bidder valuations is completely
solved. As was mentioned before, a $(1-1/e)$-approximation algorithm
for the problem exists \cite{V08}.

 %%%%%%%%%%%%%%%%%%%%%%%%%%%%%%%%%%%%%%%%%%%%%%%%%%%%%%%%%%%%
 \section{Applications of submodular optimization}
 
 %\begin{comment}
 As was stated before, submodular optimization has applications in other
problems and areas. In this section we give some examples of such applications.
In \ref{app:1}, we consider applications of submodular optimization in social
network problems in which the most influential subgroup of a society is of our
concern. In \ref{app:2} we analyze the problem of finding correspondent words
in translated sentences  called \emph{word alignment}. In \ref{app:3}, we want
to summarize a number of related documents. In all of these problem, an
optimization problem is introduced for which the correspondent objective
function is submodular and efficient maximization methods for this class of
 functions are used to solve the problem.
%\end{comment}

 \subsection{Social Network\label{app:1}}
Assume you have a product and you want to advertise it in Facebook, but you
have a limited budget and hence you can present your advertisement to a limited
number of Facebook users. Indeed, some users are more social and effective in
the society while some others are isolated. 
Social users can help distribute the information while the capability of
isolated individuals for doing so is small. 
Therefore it is reasonable to
present your product to the most influential nodes of the society.
Domninigos and Richardson first
posed this problem as a fundamental algorithmic problem.
The
optimization problem of finding such nodes is NP-hard in general, however
acceptable approximation solutions exist for the problem. Using an analysis
framework based on submodular functions, it is shown that a natural greedy
algorithm, one can find a solution which is within $63\%$ of the optimal
solution of the problem \cite{KKT03}.

 \subsection{Word Alignment \label{app:2}}
 Assume that we have a sentence in English as well as its French translation
and we want to see the correspondence of the words in the two sentences so that
we can know which French words correspond to each English word in the sentence.
In general, we can model any correspondence between the two sentences by a
bipartite graph. The nodes in the first part are the words in English while the 
nodes in the second part are the words in French. Each bipartite graph is
uniquely determined by the set of its edges (say $A$) which is a subset of the
collection of all the edges (say $V$). Assume that we have a function $f$ which
measures the quality of a correspondence $A\subset V$ as a real nonnegative
number. The words alignment problem is equivalent to maximizing $f: 2^V
\rightarrow \mathbb{R}_{+}$ under certain constraints. When $f$ is monotone and
submodular, near-optimal solutions for this problem exit \cite{BL11}.

 \subsection{Document summarization\label{app:3}}
 Assume we have a collection of related documents which we want to summarize.
The way to approach this problem is to define appropriate objective functions
and optimization problems. This problem is called \emph{multi-document
summarization}. A number of appropriate objective functions could be found in
\cite{CG98, FV04, TO09, RFH10, SL10}. It is seen that these well-established
summarization methods correspond to submodular function optimization
\cite{BL11}. Therefore, simple greedy algorithms for monotone submodular
function maximization could be used for this problem to guarantee a
summarization which is almost as good as the best possible solution obtained by
explicitly solving the optimization problem.

%  \cite{LB11}.
%  
%  In this paper, we address the problem of generic and query-based extractive summarization from collections of related documents, a task commonly known as \textit{multi-document summarization}. We treat this task as monotone submodular function maximization. This has a number of critical benefits.
% \begin{itemize}
%  \item there exists a simple greedy algorithm for monotone submodular function maximization where the summary solution obtained is guaranteed to be almost as good as the best possible solution according to an objective F.
%  \item Of course, none of this is useful if the objective function F is inappropriate for the summarization task. In this paper, we argue that monotone nondecreasing submodular functions F are an ideal class of functions to investigate for document summarization. We show, in fact, that many well-established methods for summarization (Carbonell and Goldstein, 1998; Filatova and Hatzivassiloglou, 2004; Takamura and Okumura, 2009; Riedhammer et al., 2010; Shen and Li, 2010) correspond to submodular function opti- mization, a property not explicitly mentioned in these publications.
% 
%  \end{itemize}

 %\begin{itemize}
 %\item Social Network.
 %\begin{itemize}
 %\item

 %\end{itemize}
 %\item 
 %Document summarization, Word Alignment.
 %\begin{itemize}
 %\item 
 %A Class of Submodular Functions for Document Summarization \nocite{}[LB11]
 %\item 
 %Word Alignment via Submodular Maximization over Matroids. \nocite{}[BL11]
 %\end{itemize}
 %\item 
 %Machine learning:
 %\begin{itemize}
 %\item ?
 %\end{itemize}
 %\end{itemize}

\bibliographystyle{alpha}
 \bibliography{refs.bib}

 \end{document}